\documentclass[%
 aip,
 amsmath,amssymb,
 reprint,%
]{revtex4-1}

\usepackage{graphicx}
\usepackage{dcolumn}
\usepackage{bm}

\usepackage[utf8]{inputenc}
\usepackage[T1]{fontenc}
\usepackage{mathptmx}
\usepackage{etoolbox}
\usepackage[separate-uncertainty=true]{siunitx}
\usepackage{comment}
\usepackage{float}

\makeatletter
\def\@email#1#2{%
 \endgroup
 \patchcmd{\titleblock@produce}
  {\frontmatter@RRAPformat}
  {\frontmatter@RRAPformat{\produce@RRAP{*#1\href{mailto:#2}{#2}}}\frontmatter@RRAPformat}
  {}{}
}%
\makeatother
\begin{document}


\title[Scaling of thin wire cylindrical compression after 100 fs Joule surface heating with material, diameter and laser energy]{Scaling of thin wire cylindrical compression after 100 fs Joule surface heating with material, diameter and laser energy}
\author{L. Yang*}
\email{yanglong@hzdr.de}
\affiliation{Helmholtz-Zentrum Dresden - Rossendorf, Bautzner Landstraße 400, 01328 Dresden, Germany}
\affiliation{Technische Universit\"at Dresden, 01062 Dresden, Germany}

\author{M.-L. Herbert}%
\affiliation{Helmholtz-Zentrum Dresden - Rossendorf, Bautzner Landstraße 400, 01328 Dresden, Germany}
\affiliation{Universit\"at Rostock, Institut für Physik, D-18051 Rostock, Germany}
\author{C. B\"ahtz}
\affiliation{Helmholtz-Zentrum Dresden - Rossendorf, Bautzner Landstraße 400, 01328 Dresden, Germany}
\author{V. Bouffetier}
\affiliation{Helmholtz-Zentrum Dresden - Rossendorf, Bautzner Landstraße 400, 01328 Dresden, Germany}
\author{E. Brambrink}
\affiliation{European XFEL, Holzkoppel 4, 22869, Schenfeld, Germany}
\author{T. Dornheim}
\affiliation{Center for Advanced Systems Understanding (CASUS), D-02826 G\"orlitz, Germany}
\affiliation{Helmholtz-Zentrum Dresden - Rossendorf, Bautzner Landstraße 400, 01328 Dresden, Germany}
\author{N. Fefeu}
\affiliation{CNRS - Université de Bordeaux, CELIA, UMR 5107, 43 rue Pierre Noailles 33405 Talence, France}
\author{T. Gawne}
\affiliation{Center for Advanced Systems Understanding (CASUS), D-02826 G\"orlitz, Germany}
\affiliation{Helmholtz-Zentrum Dresden - Rossendorf, Bautzner Landstraße 400, 01328 Dresden, Germany}
\author{S. G\"ode}
\affiliation{European XFEL, Holzkoppel 4, 22869, Schenfeld, Germany}
\author{J. Hagemann}
\affiliation{DESY Deutsches Elektronen-Synchrotron, Photon Science, Notkestrasse 85 - 22607 Hamburg, Germany}
\author{H. H\"oeppner}
\affiliation{Helmholtz-Zentrum Dresden - Rossendorf, Bautzner Landstraße 400, 01328 Dresden, Germany}
\author{L. G. Huang*}
\affiliation{Helmholtz-Zentrum Dresden - Rossendorf, Bautzner Landstraße 400, 01328 Dresden, Germany}
\email{lingen.huang@hzdr.de}
\author{O. S. Humphries}
\affiliation{European XFEL, Holzkoppel 4, 22869, Schenfeld, Germany}
\author{T. Kluge}
\affiliation{Helmholtz-Zentrum Dresden - Rossendorf, Bautzner Landstraße 400, 01328 Dresden, Germany}
\author{D. Kraus}
\affiliation{Universit\"at Rostock, Institut für Physik, D-18051 Rostock, Germany}
\affiliation{Helmholtz-Zentrum Dresden - Rossendorf, Bautzner Landstraße 400, 01328 Dresden, Germany}
\author{J. L\"utgert}
\affiliation{Universit\"at Rostock, Institut für Physik, D-18051 Rostock, Germany}
\author{J.-P. Naedler}
\affiliation{Universit\"at Rostock, Institut für Physik, D-18051 Rostock, Germany}
\author{M. Nakatsutsumi}
\affiliation{European XFEL, Holzkoppel 4, 22869, Schenfeld, Germany}
\author{A. Pelka}
\affiliation{Helmholtz-Zentrum Dresden - Rossendorf, Bautzner Landstraße 400, 01328 Dresden, Germany}
\author{T. R. Preston}
\affiliation{European XFEL, Holzkoppel 4, 22869, Schenfeld, Germany}
\author{C. Qu}
\affiliation{Universit\"at Rostock, Institut für Physik, D-18051 Rostock, Germany}
\author{S. V. Rahul}
\affiliation{European XFEL, Holzkoppel 4, 22869, Schenfeld, Germany}
\author{R. Redmer}
\affiliation{Universit\"at Rostock, Institut für Physik, D-18051 Rostock, Germany}
\author{M. Rehwald}
\affiliation{Helmholtz-Zentrum Dresden - Rossendorf, Bautzner Landstraße 400, 01328 Dresden, Germany}
\author{L. Randolph }
\affiliation{European XFEL, Holzkoppel 4, 22869, Schenfeld, Germany}
\author{J. J. Santos}
\affiliation{CNRS - Université de Bordeaux, CELIA, UMR 5107, 43 rue Pierre Noailles 33405 Talence, France}
\author{M. \v{S}m\'{\i}d}
\affiliation{Helmholtz-Zentrum Dresden - Rossendorf, Bautzner Landstraße 400, 01328 Dresden, Germany}
\author{U. Schramm}
\affiliation{Helmholtz-Zentrum Dresden - Rossendorf, Bautzner Landstraße 400, 01328 Dresden, Germany}

\affiliation{Technische Universit\"at Dresden, 01062 Dresden, Germany}

\author{J.-P. Schwinkendorf  }
\affiliation{European XFEL, Holzkoppel 4, 22869, Schenfeld, Germany}
\affiliation{Helmholtz-Zentrum Dresden - Rossendorf, Bautzner Landstraße 400, 01328 Dresden, Germany}
\author{M. Vescovi}
\affiliation{Helmholtz-Zentrum Dresden - Rossendorf, Bautzner Landstraße 400, 01328 Dresden, Germany}

\author{U. Zastrau}
\affiliation{European XFEL, Holzkoppel 4, 22869, Schenfeld, Germany}
\author{K. Zeil}
\affiliation{Helmholtz-Zentrum Dresden - Rossendorf, Bautzner Landstraße 400, 01328 Dresden, Germany}

\author{A. Laso Garcia*}
\affiliation{Helmholtz-Zentrum Dresden - Rossendorf, Bautzner Landstraße 400, 01328 Dresden, Germany}
\email{a.garcia@hzdr.de}

\author{T. Toncian}
\affiliation{Helmholtz-Zentrum Dresden - Rossendorf, Bautzner Landstraße 400, 01328 Dresden, Germany}
\author{T. E. Cowan}
\affiliation{Helmholtz-Zentrum Dresden - Rossendorf, Bautzner Landstraße 400, 01328 Dresden, Germany}
\affiliation{Technische Universit\"at Dresden, 01062 Dresden, Germany}

\date{\today}

\begin{abstract}
We present the first systematic experimental validation of return-current-driven implosion scaling in micrometer-sized wires irradiated by femtosecond laser pulses. Employing XFEL-based imaging with sub-micrometer spatial and femtosecond temporal resolution, supported by hydrodynamic and particle-in-cell simulations, we reveal how return current density depends precisely on wire diameter, material properties, and incident laser energy. We identify deviations from simple theoretical predictions due to geometrically influenced electron escape dynamics. These results refine and confirm the scaling laws essential for predictive modeling in high-energy-density physics and inertial fusion research.

\end{abstract}

\maketitle

\section{\label{sec:level1} Introduction}
Achieving controlled fusion energy in the laboratory remains one of the most profound challenges in modern physics. At the heart of inertial confinement fusion (ICF) research lies the quest to compress hydrogenic fuel to extreme densities and temperatures, a regime recently accessed by large-scale, multi-kilojoule, and nanosecond-pulse laser facilities such as the National Ignition Facility (NIF) and OMEGA~\citep{zylstra2022burning,williams2024demonstration,gopalaswamy2024demonstration}. These flagship experiments have, for the first time, demonstrated net energy gain and enabled exploration of matter at multi-gigabar pressures. However, the high cost, low repetition rate, and diagnostic limitations of such facilities impede systematic investigation of the complex, transient plasma dynamics underlying heating, compression, and instability growth~\citep{betti2016inertial,lindl2004physics,glenzer2009x,hurricane2014fuel}. Conventional optical and X-ray backlighting techniques are challenged by the high opacity and ultrafast evolution of the compressed plasma core, leaving important questions unresolved regarding energy coupling, hydrodynamic instability, and material response at relevant pressure, spatial, and temporal scales~\citep{lindl2004physics,moses2009national,glenzer2009x,hurricane2014fuel,betti2016inertial}.

Recent studies have shown that short-pulse lasers operating at the joule scale can induce micrometer-scale cylindrical implosions, achieving pressures approaching \SI{1}{Gbar} at stagnation based on the hydrodynamic simulation predictions~\citep{laso2024cylindrical,yang2024dynamic}. These implosions, driven by surface ablation initiated by hot-electron-induced return currents, provide a highly controllable platform for investigating fundamental high-energy-density (HED) plasma processes. When combined with femtosecond hard X-ray free-electron laser (XFEL) diagnostics, these experiments enable unprecedented temporal (fs) and spatial (sub-micrometer) resolution, directly revealing processes previously inaccessible or unclear, such as instability growth and decay \cite{ordyna2024visualizing}, hole-boring dynamics in wire targets \cite{Nanoscale2016}, and laser-driven fast electron transport in solid-density matter \cite{2023Ultrafast,huang2025demonstration}. Such capability opens new windows into fundamental questions of implosion dynamics and equation-of-state behavior at extremely high pressures.

\begin{figure*}
\includegraphics[width=1\textwidth]{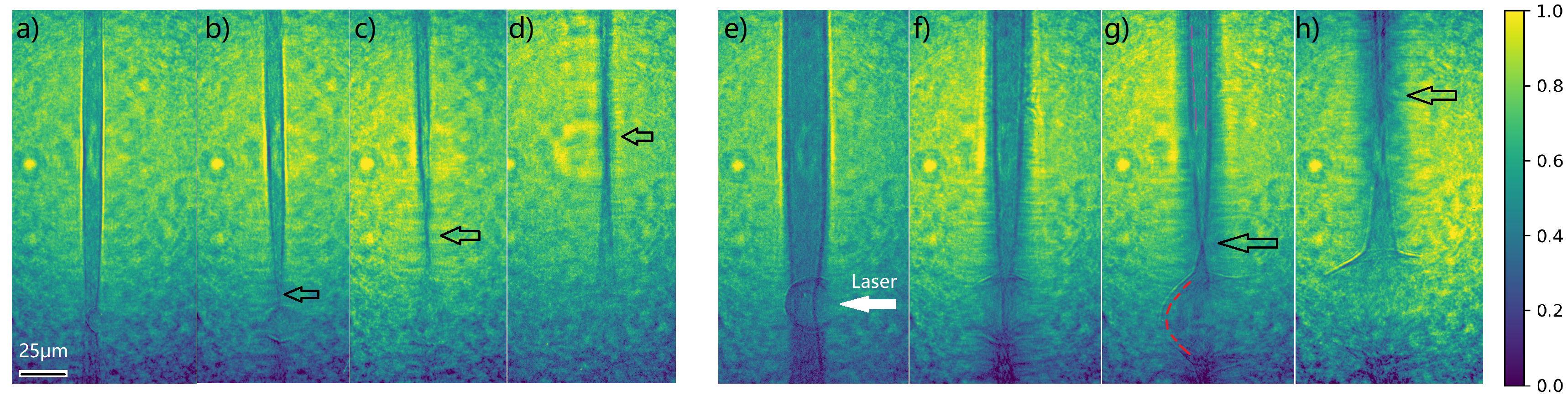}
\caption{X-ray images of wire implosions at different time delays. Panels (a)-(d) show a \SI{10}{\um} diameter copper wire at delays of 40 ps, 100 ps, 200 ps, and 300 ps after the ReLaX laser. Panels (e)-(h) show a \SI{25}{\um} diameter copper wire at delays of 250 ps, 500 ps, 700 ps, and 1000 ps, respectively. The color scale represents the square root of normalized X-ray photon density on the Zyla detector. The ReLaX laser is incident from the right, as indicated by the white arrow in panel (e). The implosion event is indicated by the hollow black arrow in each panel. In panel (g), examples of a blast shock wave at the laser focus and a cylindrical compression shock wave on the wire surface are illustrated by the red and pink dashed curves, respectively. }
\label{fig:fig2_data}
\end{figure*}

The physical mechanism driving such implosions involves surface-bound return currents triggered by laser-accelerated hot electrons. When an intense fs laser irradiates a solid target, electrons with energy exceeding MeV are expelled and leave behind a net positive charge\citep{gibbon1994efficient,malka1996experimental,gibbon2005short}; rapid neutralization by return currents flowing along the target surface \cite{snavely2000intense,kaymak2016nanoscale,beg2004return,hauer1983return,benjamin1979measurement} can, under the right conditions, drive compressive or pinching forces in wire‐like geometries.  Indirect evidence of this process was first observed in solid hydrogen‐jet targets via optical shadowgraphy probing on the DRACO laser \citep{Martin2022,yang2024dynamic}. Most notably, the Europe XFEL diagnostics have recently enabled the first direct observation of cylindrical implosions in micrometer-sized metallic wires \citep{laso2024cylindrical}. While the proof‐of‐principle study demonstrates the promise of combining fs lasers with XFEL diagnostics for a new path to the highest pressures relevant to ICF research, it remains limited to a single‐case demonstration.

A systematic understanding of return‐current implosion requires quantitative validation of the underlying scaling laws—specifically, how the peak surface current density \(j\) depends on (i) wire radius \(r\), (ii) atomic number \(Z\), and (iii) incident laser energy \(E_L\). Theoretically, return‐current models predict an inverse‐radius scaling \(j\propto r^{-1}\) with only weak material dependence, modified by hot‐electron escape dynamics and surface‐wave attenuation \citep{yang2024dynamic}. Experimental confirmation of these predictions across a broad parameter space is essential both for refining our physical models and for guiding the design of next‐generation relativistic fs‐laser implosion experiments.

In this work, we present the first comprehensive experimental validation of return‐current scaling in micrometer‐sized wires. By combining XFEL‐backlit imaging (with sub-micrometer spatial and femtosecond temporal resolution) with hydrodynamic and particle‐in‐cell (PIC) simulations, we quantify \(j\) over a range of wire diameters (\SI{10}{\um}–\SI{25}{\um}), materials (Cu vs.\ Al), and laser energies (0.23–3 J). We show that \(j\) indeed scales inversely with \(r\), exhibits minimal \(Z\) dependence, and follows the predicted \(E_L^{2/3}\) law, with systematic deviations of \SI{5}{\%} captured by geometry‐ and attenuation‐based correction factors. Our results not only bridge critical gaps between theory and observation but also establish the scaling behavior of return-current-induced implosions and demonstrate the predictive capability of the XFEL + fs-laser platform for studying ICF-relevant high pressure physics at reduced scale.

\section{\label{sec:level2} Experimental Setup}

The experiment (p5689) was executed at the HED-HiBEF instrument \cite{zastrau2021high} at the European X-ray Free Electron Laser, utilizing the ReLaX optical laser system (maximum power 300\,TW) as the pump source. The ReLaX laser (wavelength of 800\,nm) irradiated the targets at a 45$^\circ$ angle relative to the X-ray propagation axis. The pulse energy reaching the target was of \SIrange{0.6}{3}{J}, with a pulse duration under 30\,fs. The focus on target had a full-width half-maximum (FWHM) of \SI{3}{\um}. The resulting maximum laser intensity was $\sim 10^{20}$\,W/cm$^2$. The 8.5\,keV X-rays generated by the SASE2 undulator were used to illuminate a square region (\SI{250}{\um})$^2$ around the ReLaX focal spot. 

The target plane was imaged via a compound refractive lense stack of 15 beryllium lenses. Each lens had a radius of curvature of \SI{50}{\um} with a web thickness of \SI{30}{\um}. The focal length of the lens stack was 36\,cm. The distance from the target to the imaging detector was 631\,cm. The detector was a GAGG scintillator imaged to an Andor Zyla CMOS camera via either a 7.5$\times$ or a 2$\times$ objective. The detector pixel pitch is  \SI{6.5}{\um}. After accounting for the total magnification factor, it results in an equivalent pixel size on target of 56\,nm/pixel for the 7.5$\times$ objective and 210\,nm/pixel for the 2$\times$ objective. 

The X-ray energy was characterized via elastic scattering on a Kapton foil with a von Hamos spectrometer. The pulse energy was $\sim\SI{700}{\uJ}$, measured with an X-ray gas monitor in the tunnel. Imaging resolution was determined using a resolution test target (NTT-XRESO 50HC). Data were collected in the self-seeded mode of the X-ray, which provides improved spectral coherence and a narrow bandwidth ($\sim < \SI{1}{eV}$), resulting in a measured imaging resolution better than 200\,nm.

\begin{figure*}[htb]
\centering 
\includegraphics[width=0.95\textwidth]{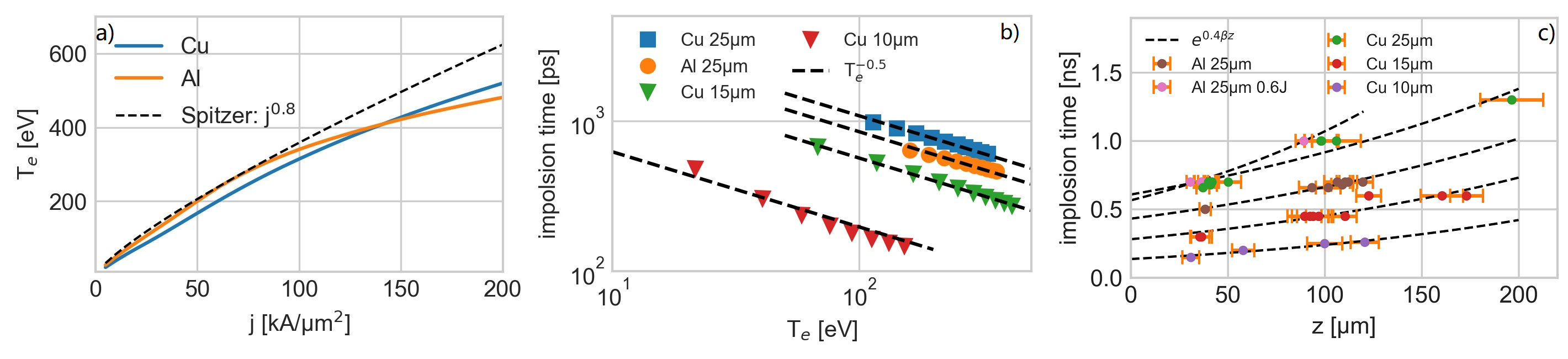} 
\caption{(a) Target surface temperature varied with peak current density. The dashed lines show the temperature 
 current density scaling using Spitzer resistivity model. (b) Target implosion time varied with different surface temperatures obtained from 1D hydrodynamic simulations. The dashed lines show the strong shock wave scaling. (c) Target implosion time along the wire \textit{z} position obtained from experiment measurements. The dashed lines show an exponential fit of the data. The default laser energy is 3 J.}
\label{fig:z_im} 
\end{figure*}

The delay between the optical pump and the X-ray probe was scanned from a few picoseconds up to nanoseconds. This allows to image the convergence of the implosion at different positions along the wire axis. A representative dataset is shown in figure~\ref{fig:fig2_data}, where the implosion dynamics of \SI{10}{\um} and \SI{25}{\um} copper wire are recorded at various time delays. Both the blast shock wave generated by the laser focal hotspot and the cylindrical shock wave emerging at the wire surface and compressing inward are clearly observed, as illustrated in figure~\ref{fig:fig2_data}(g). Implosion events resulting from cylindrical compression are indicated by hollow black arrows in each panel.

\section{Return current scaling law validation}

\subsection{Current density distribution on the wire surface}

In Ref.~\citep{yang2024dynamic}, it was demonstrated that electrons escaping from an intense laser-irradiated target induce a return current flowing along the wire's surface. Due to the ultrashort ($\sim \SI{100}{fs}$) duration of the current pulse, the return current is confined within the skin-depth layer of the target surface. The transient current rapidly heats this thin surface plasma layer to temperatures of several hundred electronvolts (eV) within a 100 femtosecond timescale, while maintaining solid densities. The subsequent rapid ablation generates inward-propagating shocks, driving a cylindrical shock compression of the wire target. Given the extremely short duration of the current pulse, the surface plasma temperature can be treated as an initial condition for the subsequent hydrodynamic evolution. Consequently, a direct correspondence exists between the surface temperature, the current density, and the resulting implosion time. Building upon these relationships, we derive that the surface current density on a thin wire satisfies the return‐current scaling law
\begin{equation}
  \frac{j_2}{j_1} \approx \frac{r_1}{r_2}\,,
  \label{eq:scaling}
\end{equation}
where $j_i$ and $r_i$ are the current density and radius for two different wire configurations. To verify Eq.~\eqref{eq:scaling} experimentally, it is necessary to reconstruct the spatial profile \( j(z) \), where \( z \) is the distance along the wire axis from the laser focal spot. However, direct measurement of \( j(z) \) is not feasible. We therefore employ a cascade of mappings
\begin{align}
  T_e &= f_1\bigl(j\bigr)\,,\label{eq:f1}\\
  \tau_{\rm im} &= f_2\bigl(T_e\bigr)\,,\label{eq:f2}\\
  j(z) &= f_3\bigl(z,\tau_{\rm im}\bigr)\,,\label{eq:f3}
\end{align}
where $f_1$ is obtained from the electron‐energy equation relating current density $j$ to electron temperature $T_e$, $f_2$ is derived via hydrodynamic simulations mapping $T_e$ to implosion time $\tau_{\rm im}$, and $f_3$ is determined from experimental measurements of the implosion time along $z$ as shown in figure \ref{fig:fig2_data}.  Combining Eqs.~\eqref{eq:f1}–\eqref{eq:f3} yields
\begin{equation}
  \label{eq:fzj}
  j(z)= f_3\bigl(z,\,f_2\bigl(f_1(j)\bigr)\bigr)\,,
\end{equation}
which provides the reconstructed surface‐current distribution as a function of \textit{z} position.\\

It is shown that return currents have two main consequences to the wire target \cite{yang2024dynamic}, one is the Z-pinch effect, another is the joule heating effect. In the wire targets we present (\SIrange[]{10}{25}{\um} diameters), the first effect can be ignored, as the magnetic pressure is much smaller than the ablation pressure. Thus, we can use the surface temperature to indirectly calculate the surface return current. In order to construct the mapping \(f_1\colon j\mapsto T_e\), we solve the electron‐energy equation \citep{yang2024dynamic}:
\begin{equation}
\frac{3}{2}n_e\frac{\partial T_e}{\partial t}
=\frac{1}{r}\frac{\partial}{\partial r}\!\Bigl(r\,K_{T_e}\frac{\partial T_e}{\partial r}\Bigr)
+\frac{j(r)^2}{\sigma_{T_e}}\;,
\label{eq:heat_transfer}
\end{equation}
where \(K_{T_e}\) and \(\sigma_{T_e}=n_e e^2/(m_e\nu_{ei})\) are the electron thermal and electrical conductivities, respectively. For temperatures lower than \SI{100}{eV}, the Burgess model is used \cite{burgess1986electrical}, for temperatures higher than \SI{100}{eV}, the SESAME database is used \cite{johnson1994sesame}.

The full numerical solutions of equation \eqref{eq:heat_transfer} are shown in figure~\ref{fig:z_im}(a) for copper and aluminum ($d$ = \SI{25}{\um}) wires. For copper wires, the results for wire diameters of 10, 15, and \SI{25}{\um} are identical and thus represented by a single curve. The weak radius dependence further confirms that Joule heating and heat transfer are limited in a very thin layer. The simulations cover peak current densities ranging from \SI{5}{kA/\um^2} to \SI{200}{kA/\um^2}, with further details provided in Appendix~A. As a guide, the Spitzer‐resistivity \cite{richardson20192019} scaling without radial heat transfer
\begin{equation}
T_e \propto j^{0.8},
\label{eq:te-j}
\end{equation}
is plotted as a dashed line in figure~\ref{fig:z_im}(a).
It can be seen that there are modest deviations for those two temperatures at \(j<50\,\mathrm{kA/\mu m^2}\) (where Spitzer overestimates the collision rate) and at \(j>100\,\mathrm{kA/\mu m^2}\) (where radial heat transfer loss becomes significant).

Next, the mapping \(f_2\colon T_e\mapsto \tau_{\rm im}\) is obtained from one‐dimensional hydrodynamic simulations (see Appendix A and Ref. \cite{yang2024dynamic,laso2024cylindrical}).  In the ablation‐driven, strong‐shock limit (over $\sim \SI{100}{Mbar}$ and $ \sim \SI{100}{eV}$ in the skin depth layer, see figures \ref{fig:z_im}(a), (b)), the Rankine–Hugoniot relation results a shock velocity, which scales as \(U_s\propto T_e^{1/2}\). As an approximation for small radius targets, the average shock velocity $\overline{U_s} \propto U_s$, so that
\begin{equation}
\tau_{\rm im}
=\frac{r_0}{\overline{U_s}}\propto T_e^{-1/2}.
\label{eq:tau-te}
\end{equation}
Figure~\ref{fig:z_im}(b) shows \(\tau_{\rm im}(T_e)\) for different wire geometries, with dashed lines indicating the \(T_e^{-0.5}\) fit. All four cases exhibit a strong agreement between the fitting and the simulation data. The simulations suggest that the implosion process is a strong shock process driven by the surface ablation. Thus, the initial surface temperature distribution after surface return current heating along the wire position \textit{z} can be obtained by incorporating the experimentally measured implosion time $\tau_{im}$.\\

\begin{figure}[H]
\centering 
\includegraphics[width=0.5\textwidth]{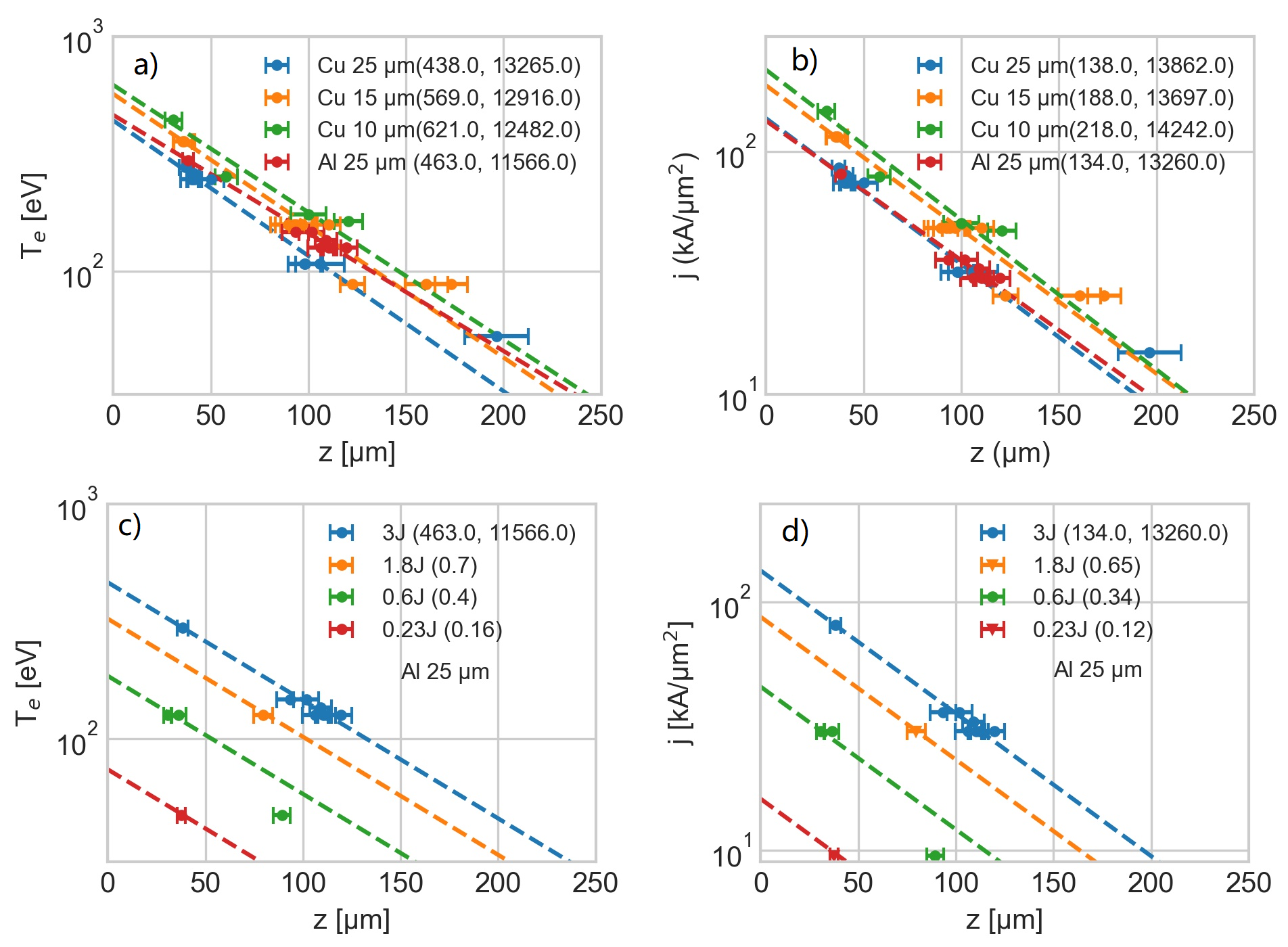} 
\caption{Panels (a) and (c): reconstructed initial surface temperature. Panels (b) and (d): reconstructed peak surface current density \( j \) along the wire axis \( z \). Dotted data points with error bars represent values obtained from figure~\ref{fig:z_im}. Dashed lines show exponential fits \( y(z) = \alpha \exp(-\beta z) \) to the reconstructed data. The two numbers in parentheses in each figure label denote the fitting constants: \(\alpha\) (units of eV for panels (a) and (c), and \SI{}{kA/\um^{2}} for panels (b) and (d)) and \(\beta\) (units of \SI{}{m^{-1}}), respectively. Panels (c) and (d) show results rescaled to the 3\,J, \SI{25}{\um} Al wire reference case; in these panels, the single number in parentheses is the normalization factor.}
\label{fig:scaling} 
\end{figure}

Finally, the mapping \(f_3\colon (z,\tau_{\rm im})\mapsto j(z)\) is extracted from experimental measurements of implosion time versus axial position (assuming the laser focus is \textit{z}= \SI{0}{\um})  as shown in figure~\ref{fig:z_im}(c). Details of the experimental data extraction procedure and the derivation of error bars are given in Appendix~B. Due to the recirculation of hot electrons~\cite{snavely2000intense,hatchett2000electron,cowan2004ultralow,Huang2022,rehwald2023ultra}, target bulk displaced from the laser focal spot can experience bulk heating driven by return currents induced by laser-accelerated hot electrons~\cite{yang2023time}. Two-dimensional and three-dimensional PIC simulations, accounting for known temperature overestimation effects in the PIC simulations~\cite{yang2023time,yang2024dynamic,yang2024thesis}, indicate that significant bulk return-current heating can extend up to approximately \( z \sim \SI{30}{\um} \) from the laser focus. This bulk heating increases the local sound speed and can lead to an overestimation of the surface return current. Further details on the selection criteria are provided in Appendix~C. To mitigate this effect, we only consider implosion data acquired from regions located beyond \SI{30}{\um} from the focal spot, and thus assume a cold initial state for the analysis presented in this work. As the return current propagated along the wire surface with surface wave behavior, the current density along the \textit{z} position of the wire propagates with \cite{yang2024thesis,yang2025} 
\begin{equation}
\label{eq:j-z}
    j(z,t)\propto j_0\bigl(z - v_g t\bigr)\,\exp\bigl[-\beta\,z\bigr],
\end{equation}
where $j_0$ is the initial current density at $z=0$, $\beta$ is the decay constant associated with wire impedance, and $v_g$ is the group velocity of the surface current wave. By incorporating equations \eqref{eq:te-j}, \eqref{eq:tau-te} and \eqref{eq:j-z}, it has
\begin{equation}
\tau_{\rm im} (z)\propto\exp\bigl[0.4\,\beta\,z\bigr],
\label{eq:im-z}
\end{equation}
showing that the implosion time distribution $\tau_{im} (z)$ is exponentially increasing with \textit{z} position. The dashed lines in figure \ref{fig:z_im}(c) represent the fitting of the experimental data using equation \eqref{eq:im-z}. A strong agreement between the experimental data and the fitting confirms the surface wave behavior on the wire targets. Because the decay constant \( \beta \) represents impedance attenuation along the wire surface, it depends primarily on geometry, material properties, and wave frequency—conditions that were similar across all our experimental cases. Consequently, all cases exhibit very similar decay constants, as illustrated in figure ~\ref{fig:z_im}(c). This characteristic enables the reconstruction of the current density distribution on the target surface using limited experimental data, making a systematic validation of the scaling law feasible. By composing these three mappings as shown in Eq.~\eqref{eq:fzj}, we reconstruct the initial surface temperature and the peak surface current density profile along the wire as shown in figure~\ref{fig:scaling}. Having reconstructed the current-density profile along the wire, we next examine how this profile’s characteristics scale with wire radius, material, and laser energy.\\

\subsection{Current‐density radius and atomic‐number dependence}

Figures~\ref{fig:scaling}(a) and (b) present the reconstructed initial surface temperature and peak current density distributions for copper and aluminum wires of various radii irradiated by a \SI{3}{J} laser (shown as dotted data). Figures~\ref{fig:scaling}(c) and (d) show corresponding results for aluminum wires of fixed diameter (\SI{25}{\um}) under varying laser energies. The dashed curves represent exponential fits based on the surface-wave model described by Eq.~\eqref{eq:j-z}. For the radius-scaling cases, separate fitting constants are used for each curve, as indicated in the figure labels. For the energy-scaling cases, the curve corresponding to the \SI{3}{J}, \SI{25}{\um} aluminum wire is used as a baseline and linearly scaled to other laser energies. As shown in figures~\ref{fig:scaling}(a) and (b), the experimental data largely follow the fitted exponential profiles within the first \SI{100}{\um} of axial position. Beyond this range, the measured values tend to slightly exceed the model predictions, with noticeably larger uncertainties. This discrepancy arises because the Spitzer resistivity model deviates from its $T^{-3/2}$ scaling at lower temperatures (below \SI{100}{eV} in our study). Overall, the agreement between experiment and theory supports the exponential attenuation behavior described in Section~A.

Using the \SI{25}{\um} Cu wire as reference, figure~\ref{fig:scaling}(b) shows that for smaller-radius Cu wires (\SI{15}{\um} and \SI{10}{\um}), the current density at the same position increases as the wire radius decreases, consistent with the expected scaling law trend. However, when considering the actual ratio, the radius‐scaling law predicts \(r_{25}/r_{15}\approx1.67\),
whereas the experiment gives \(j_{15}/j_{25}\approx1.48\) (\(\sim19\%\) lower), and for \SI{10}{\micro\metre} wires it predicts \(r_{25}/r_{10}\approx2.50\) vs.\ measured 1.7 (\(\sim47\%\) lower) for \textit{z}= \SI{40}{\um} position. It is found that the nominal inverse‐radius scaling \(j\propto r^{-1}\) 
overestimate the current density when scaling from large radius to smaller radius target. The correction for the nominal scaling law can be studied at least in two aspects. First, according to the target normal sheath acceleration (TNSA) mechanism, a sheath field forms at the target surface, and only hot electrons with energies exceeding the sheath potential can escape \cite{snavely2000intense,hatchett2000electron,cowan2004ultralow,Huang2022,rehwald2023ultra}; the magnitude of the sheath field depends on the target geometry. The sheath field intensity at the rear of the target scales as \(E\propto d^{-2}\)\, for slab target \citep{roth2017ion}, implying that increased target thickness reduces hot electron recirculation and enhances hot electron escape as the wire diameter increases. Here the reduced escaping‐electron charge in thinner wires leads to an current density ratio correction \(\phi(r)=(r/ \SI{25}{\um})^{0.16}\), as confirmed by the analytic calculation and 2D PIC simulations of hydrogen jets as presented in Appendix D. Second, the fitted decay constant \(\beta\) encapsulates the combined effects of wire geometry (radius) and surface conductivity (electron temperature).  As shown in Fig.~\ref{fig:scaling}(b), \(\beta\) increases for smaller radii (\(\beta_{10}>\beta_{25}\)), reflecting the stronger attenuation of high‐frequency surface waves in thinner wires due to geometric filtering.\\
Combining those two corrections, the ratio of current densities at two radii \(r_1,r_2\) becomes
\begin{equation}
  \frac{j_{r_1}(z)}{j_{r_2}(z)}
  = \frac{r_2}{r_1}\,\phi(r_1,r_2)\,
    e^{\bigl[(\beta_{r_1}-\beta_{r_2})\,z\bigr]}
  \approx
  \Bigl(\frac{r_2}{r_1}\Bigr)^{0.84}
  e^{\bigl[(\beta_{r_1}-\beta_{r_2})\,z\bigr]},
  \label{eq:revised_scaling}
\end{equation}
where \(\phi(r_1,r_2)=(r_1/r_2)^{0.16}\). At \(z=\SI{40}{\um}\), this predicts \(j_{15}/j_{25}\approx1.51\) versus an experimental value of 1.48 ($\sim \SI{2}{\%}$ lower), and \(j_{10}/j_{25}\approx2.10\) versus 1.70 ($\sim \SI{20}{\%}$ lower). We note that \(\beta\) varies with \(z\): for the \SI{25}{\um} and  \SI{15}{\um} wires, two attenuation regimes emerge before and after \(z\approx \SI{100}{\um}\), indicating preferential damping of high-frequency components and a subsequent remaining lower-frequency components with reduced \(\beta\). In the  \SI{10}{\um} wire, this transition occurs by \(z\approx \SI{50}{\um}\), leading to an underestimation of the effective \(\beta\) and contributing to the \SI{20}{\%} deviation.  A detailed mapping of \(\beta(z)\) requires finer spatial resolution and will be pursued in future work.\\

Figure~\ref{fig:scaling}(b) compares the reconstructed \(j(z)\) for \SI{25}{\micro\metre} Cu and Al wires, yielding  \(j(z)_{\rm Cu}/j(z)_{\rm Al}\approx1\)
across the measured range.  This near‐unity ratio confirms that, under identical geometry and laser conditions, the return‐current scaling law is effectively material‐independent.  Two key factors explain this result: first, the initial amplitude \(j_0\propto Q_{\rm es}\) (Eq.~\ref{eq:Q_es}) is governed predominantly by wire geometry (radius) rather than atomic number; second, the propagation attenuation \(\beta\) depends on the surface conductivity, for which Cu and Al follow nearly identical Spitzer‐resistivity models at our experimental temperatures \cite{yang2024thesis}.  Consequently, both the amplitude and decay of the surface return current are insensitive to material choice under our conditions.\\

In summary, the surface return‐current density scales inversely with wire diameter—smaller radii produce systematically higher \(j(z)\), in quantitative agreement with the revised scaling law (equation~\eqref{eq:revised_scaling}).  By contrast, Cu and Al wires of identical diameter yield indistinguishable \(j(z)\), confirming that atomic number exerts a negligible influence on return‐current behavior under our conditions.  Building on these findings, in the next section we isolate geometric and material effects by studying a single \SI{25}{\um} Al wire under varying laser energies, thereby assessing the potential of ultrashort‐pulse lasers for fusion‐relevant pressure generation.\\

\subsection{Current density laser energy dependence}

Understanding how the surface return‐current density scales with incident laser energy is critical for extrapolating from the current 3 J ReLaX system—which approaches $\sim \SI{1}{Gbar}$ pressures (pressure at implosion stagnation based on hydro simulations) \citep{yang2024dynamic,laso2024cylindrical}—to the higher energies required for ICF-relevant high pressures.  We therefore systematically study the dependence of \(j\) on \(E_L\) (laser energy).\\

\begin{figure}[htb]
\centering 
\includegraphics[width=0.5\textwidth]{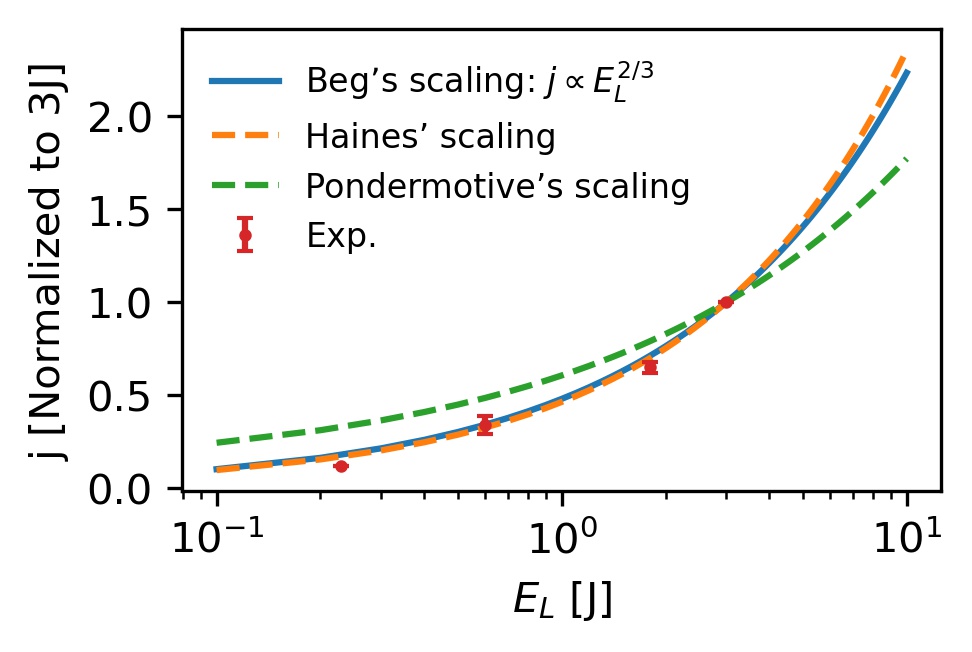} 
\caption{The surface return current density in \SI{25}{\um} Al wire varies with laser energy. Here the laser waist and pulse duration are fixed. }
\label{fig:EL_scaling} 
\end{figure} 
The dependence of the surface return current density on incident laser energy follows directly from the hot‐electron scaling.  Using \(T_h=0.469\,a_0^{2/3}\) \citep{beg1997study} and Eq.~\eqref{eq:Q_es}, one obtains
\begin{equation}
  j \;\propto\; E_L^{2/3}\,w_0^{-4/3}\,\tau_L^{-2/3}\,,
  \label{eq:j-Ew}
\end{equation}
where \(w_0\) is the laser focus waist and \(\tau_L\) is the laser pulse width. Those two parameters are fixed in our experiments.  Hence \(j\propto E_L^{2/3}\) when \(w_0\) and \(\tau_L\) are held constant.

Figure~\ref{fig:EL_scaling} plots the experimentally extracted amplitude factors \(C\) for a \SI{25}{\micro\metre} Al wire at \(E_L=3,\,1.8,\,0.6,\) and \(0.23\)\,J, namely \(C=1.00,\;0.65,\;0.34,\;0.12\) as shown in figure \ref{fig:scaling}(c)-(d).  Normalizing each to the \(3\) J case and comparing with the predicted \(\bigl(E_L/3\,\mathrm{J}\bigr)^{2/3}\) (dashed), we find close agreement across the full range. For reference, the ponderomotive scaling (energy coupling based on ponderomotive potential) and Haines’ resistive scaling are plotted for comparison \citep{wilks2001energetic,haines2009hot} (also normalized at 3 J), confirming that hot‐electron dynamics reliably predict the peak return current and, by extension, the attainable pressure as a function of laser energy.

This scaling indicates that by increasing the laser energy, one can predictably raise the peak return current (and thus implosion pressure), informing the design of next-generation high-energy experiments.

\section{\label{sec:level2} Summary and Outlook}

Our experiments have provided the first systematic validation of return-current-driven implosion scaling, demonstrating clear dependencies on wire diameter, material, and laser energy. The comprehensive study demonstrates clear scaling behaviors with wire diameter, material, and laser energy, identifying subtle deviations due to the geometrical effects of hot-electron escape. These findings bridge a critical gap between theoretical models and experimental observations. The refined scaling laws greatly enhance our ability to predict implosion dynamics, which is essential for the development of inertial confinement fusion technology and high-energy-density plasma science. By confirming these scaling laws, we provide a solid experimental foundation for designing future high-energy experiments and for benchmarking simulation codes in high-energy-density physics. \\

\section{Acknowledgments}
We thank the European XFEL in Schenefeld, Germany, and the HiBEF user consortium for the provision X-ray laser time at the HED-HIBEF (High Energy Density Science) scientific
instrument under proposal number 5689  as part of the HIBEF priority access and thank their staff for their support and the equipment provided to make this experiment possible. TD's and TG's work was partially supported by the Center for Advanced Systems Understanding (CASUS), financed by Germany’s Federal Ministry of Education and Research (BMBF) and the Saxon state government out of the State budget approved by the Saxon State Parliament. TG has received funding from the European Union's Just Transition Fund (JTF) within the project \textit{R\"ontgenlaser-Optimierung der Laserfusion} (ROLF), contract number 5086999001, co-financed by the Saxon state government out of the State budget approved by the Saxon State Parliament. FLASH  was developed in part by the DOE NNSA- and DOE Office of Science-supported Flash Center for Computational Science at the University of Chicago and the University of Rochester. The raw experimental data are available under DOI:10.22003/XFEL.EU-DATA-005689-00 upon reasonable request. Additional data supporting this study are available from the corresponding authors upon reasonable request.

Author contributions: L.Y. conceived the original idea and developed the theoretical framework. A.L.G. and T.T. led the experiment. L.Y., C.B., V.B., E.B., T.D., N.F., T.G., S.G., J.H., H.H., L.G.H., O.S.H., T.K., D.K., J.L., J.‑P.N., M.N., A.P., T.R.P., C.Q., S.V.R., R.R., M.R., L.R., J.J.S., M.Š., U.S., J.‑P.S., M.V., U.Z., K.Z., A.L.G., T.T. and T.E.C. conducted the experiment. M.‑L.H. processed the experimental data. L.G.H. performed the two‑dimensional PLCLS simulations. L.Y. carried out the remaining PIC and hydrodynamic simulations. L.Y., M.‑L.H., L.G.H., A.L.G., T.T. and T.E.C. performed the data analysis. A.L.G. and T.T. wrote the experimental section of the paper. L.Y. wrote the remaining sections. T.E.C. contributed to the concept presented in Appendix D. All authors discussed the results and reviewed the manuscript. T.E.C supervised the project.

\begin{appendix}
\section{Hydrodynamic simulations}
Hydrodynamic simulations were performed using the FLASH code (version 4.6.2), developed by the University of Rochester \cite{fryxell2000flash,dubey2009extensible}. All simulations were carried out in one-dimensional cylindrical geometry. The computational domain spans \SI{50}{\um} in radius, with the target occupying the central region and the remainder initialized as vacuum. To facilitate numerical stability, the vacuum region was filled with low-density hydrogen (\SI{1e-8}{g/cm^3}) at a temperature of \SI{1}{eV}.
Target materials include copper and aluminum, each initialized at their respective solid densities. The initial temperature profile of the target is prescribed using the electron energy equation~\eqref{eq:heat_transfer}. Figure~\ref{fig:z_im}(a) shows the calculated peak surface temperature for different current densities. In accordance with the return-current scaling theory~\cite{yang2024dynamic}, the magnetic compression associated with the $\mathbf{J} \times \mathbf{B}$ force is neglected, and the initial fluid velocity is set to zero. Reflective boundary conditions are applied at the symmetry axis, while outflow boundary conditions are used at the outer edge of the simulation domain. A self-adaptive mesh refinement scheme is employed to resolve steep gradients. Material properties are obtained from SESAME equation-of-state tables~\cite{johnson1994sesame}. The system is evolved using the one-fluid, two-temperature (ion and electron) hydrodynamic equations implemented within FLASH. The simulated implosion times corresponding to different initial surface temperatures are shown in figure~\ref{fig:z_im}(b).

\section{Obtain implosion time from experiment data}
Here the laser irradiation position and wire implosion position have to be determined by analyzing the detector images as shown in Fig.\ref{fig:fig2_data}. The position of the shock convergence point can be read-off directly while the laser irradiation position has be determined using the mean value of the edges of the circular region around the laser impact. We selected three data sets for this study. The first data set comprises copper wires with different radius (\SI{25}{\um}, \SI{15}{\um}, and \SI{10}{\um}) and are used to investigate the radius dependence of the current density. The second data set includes \SI{25}{\um} wires made of different materials, specifically copper and aluminum, to verify the atomic number (Z) dependence of the current density. The third data set consists of \SI{25}{\um} aluminum wires irradiated to varying incident laser energies (3 J as the default, 1.8 J, 0.6 J, and 0.23 J) to examine the laser energy dependence of the current density. Each data set includes repeated runs with different XFEL delays relative to the ReLaX laser irradiation time. For each effective run, an implosion point is observed on the wire. The \textit{z} position of this implosion point is measured for each corresponding delay, resulting in pairs of \textit{z} positions and implosion times. A summary of the variation in wire implosion time with \textit{z} position is presented as the dot data in figure \ref{fig:z_im}. The error bars come from the detector's resolution, inaccuracies when reading off the positions due to low contrast in some cases or the size of the relevant features on the image and from the conversion of pixels to \SI{}{\um}.

\section{Laser-accelerated hot electrons induce bulk heating of the target at positions away from the laser focal region}
\begin{figure}[htb]
\centering 
\includegraphics[width=0.5\textwidth]{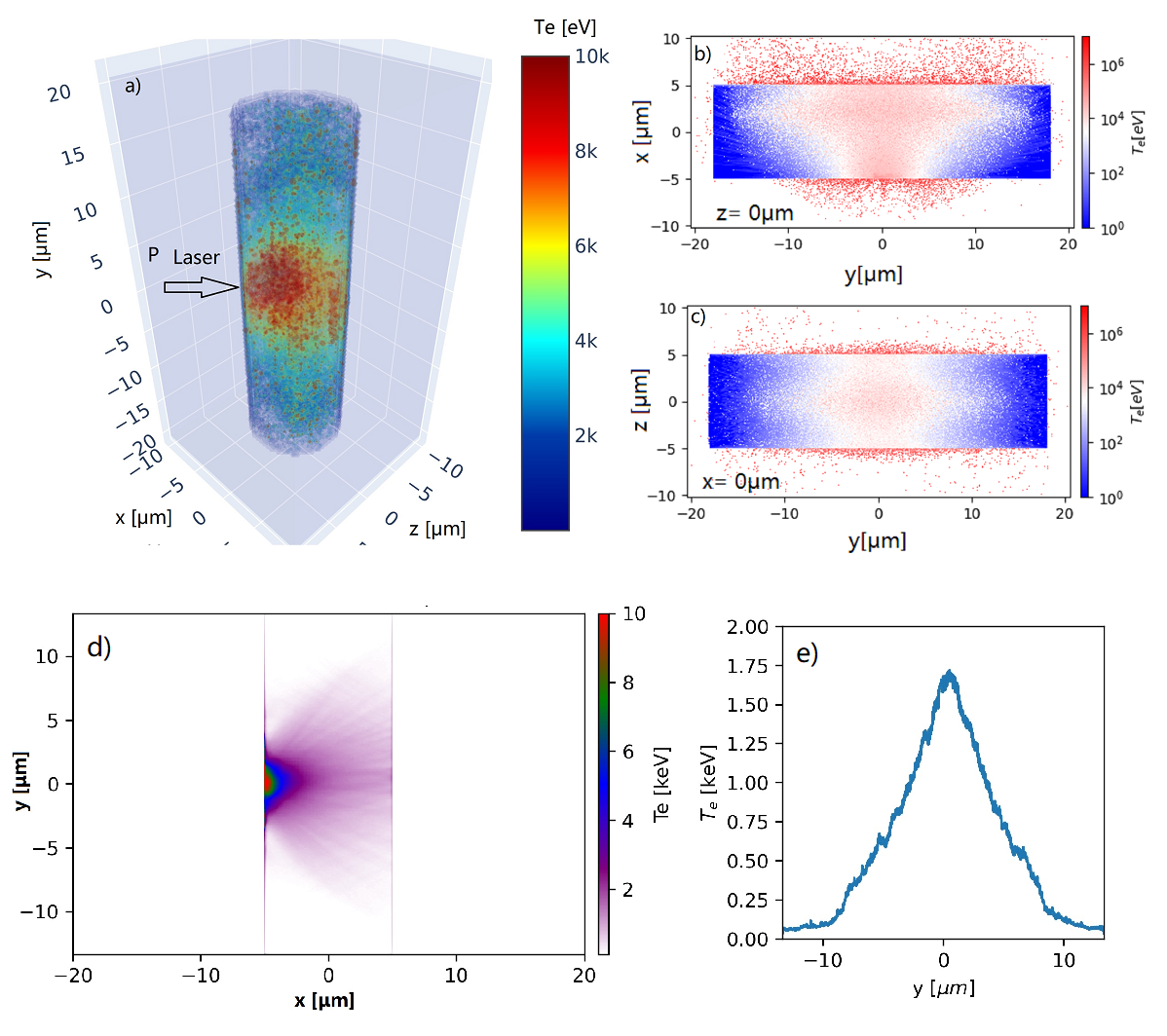} 
\caption{(a) The 3D electron temperature distribution at time of 148 fs simulated by 3D PIConGPU.
(b)-(c) The cross section of 2D electron temperature distribution at $z = \SI{0}{\um}$ and $x = \SI{0}{\um}$, respectively. (d) The copper temperature distribution at 113 fs after the laser main pulse in 2D PICLS simulations. (e) The
lineout of the temperature distribution at $x=\SI{0}{\um}$.}
\label{fig:bulk_te} 
\end{figure}

Relativistic short-pulse lasers can significantly increase the bulk temperature of targets at regions displaced from the laser focal spot. This effect has been demonstrated in shadowgraphy experiments with solid hydrogen jets irradiated by a DRACO laser at relativistic intensities (\(a_0 \approx 1\)) \cite{yang2023time}. For metallic wires irradiated by the ReLaX laser system, three-dimensional (3D) particle-in-cell (PIC) simulations show that laser-accelerated hot electrons circulate and deposit energy as far as \SI{20}{\um} away from the laser focus in a \SI{10}{\um}-diameter copper wire target, as illustrated in Fig.~\ref{fig:bulk_te}(a)-(c). Two-dimensional (2D) PIC simulations further reveal that at \SI{13}{\um} from the laser focus, the bulk electron temperature is approximately \SI{60}{eV}, as presented in Fig.~\ref{fig:bulk_te}(d)-(e). Considering known temperature overestimations inherent in PIC simulations \cite{yang2023time,yang2024dynamic}, a reduction factor of approximately 10 is applied, yielding an estimated bulk temperature of \SI{6}{eV} at \( z = \SI{13}{\um} \). Thus, a realistic estimation suggests that hot-electron-induced heating could affect the bulk temperature up to \( z = \SIrange{20}{30}{\um} \). Details of the PIC simulation setups and corresponding results are provided in Ref.~\cite{yang2024thesis}.

\section{Nonideal Effects of Escaping Charges: Dependence on Target Radius}

In this section, we discuss the nonideal effects of the escaping charges as a function of the target radius. Outside the target wire, the hot electron density is assumed to follow a Boltzmann distribution,
\begin{equation}
    f(E)=\frac{1}{k_BT_h}e^{-\frac{E}{k_BT_h}},
\end{equation}
where $T_h$ is the hot electron temperature. The total number of hot electrons is given by
\begin{equation}
    N_t=f_L\frac{E_L}{T_h},
    \label{eq:Q_es}
\end{equation}
with $f_L$ representing a scaling factor and $E_L$ the laser energy. Hot electrons with energy $E>eV_s$ ($V_s$ is the sheath potential) escape from the target. Consequently, the fraction of escaping charges is
\begin{equation}
    \eta = \frac{N_{es}}{N_t}=\int_{eV_s}^\infty f(E)dE=e^{\frac{-eV_s}{k_BT_h}},
\end{equation}
which indicates that only those hot electrons with energies exceeding the sheath potential, $V_s$, overcome the barrier and escape into the vacuum. Simultaneously, these escaping electrons contribute to an increase in the sheath potential.

The potential generated by the hot electrons in a cylindrical system can be derived from the Poisson equation in vacuum,
\begin{equation}
    \frac{1}{r}\frac{d}{dr}\Bigl(r\frac{dV}{dr}\Bigr)=\frac{en_e}{\epsilon_0}.
\label{eq:possion}
\end{equation}
To simplify the analysis, an effective capacitance, $C_{eff}$, is introduced so that the sheath potential is given by
\begin{equation}
    V_s=\frac{Q_{es}}{C_{eff}},
\end{equation}
where $Q_{es} = eN_{es}$ is the escaping charge. Thus, the previous relation becomes
\begin{equation}
    \eta = e^{\frac{-e^2N_{es}}{C_{eff}k_BT_h}}=e^{\frac{-e^2\eta N_t}{C_{eff}k_BT_h}}.
\end{equation}
The solution of this equation yields
\begin{equation}
    \eta(r)=\frac{C_{\rm eff}(r)\,(k_BT_h)^2}{f_Le^2\,E_{\rm L}}\,W\!\Bigl(\frac{f_Le^2\,E_{\rm L}}{C_{\rm eff}(r)\,(k_BT_h)^2}\Bigr),
\label{eq:Wfunction}
\end{equation}
where $W$ is the Lambert W function. Moreover, solving Equation~\eqref{eq:possion} leads to the unit capacitance of a wire target,
\begin{equation}
    C(r)=\frac{2\pi\epsilon_0 }{\ln(r_1/r)},
\end{equation}
with $r$ being the radius of the cylinder and $r_1$ the effective radius of the outer conductor. In our case, $r_1\gg r$. Experimentally, $r_1$ is on the order of \SI{}{\mm}, and in the simulations an absorption boundary condition is applied to effectively truncate $r_1$. For practical purposes, we take $r_1=\SI{1}{\mm}$.

\begin{figure}[htb]
\centering 
\includegraphics[width=0.5\textwidth]{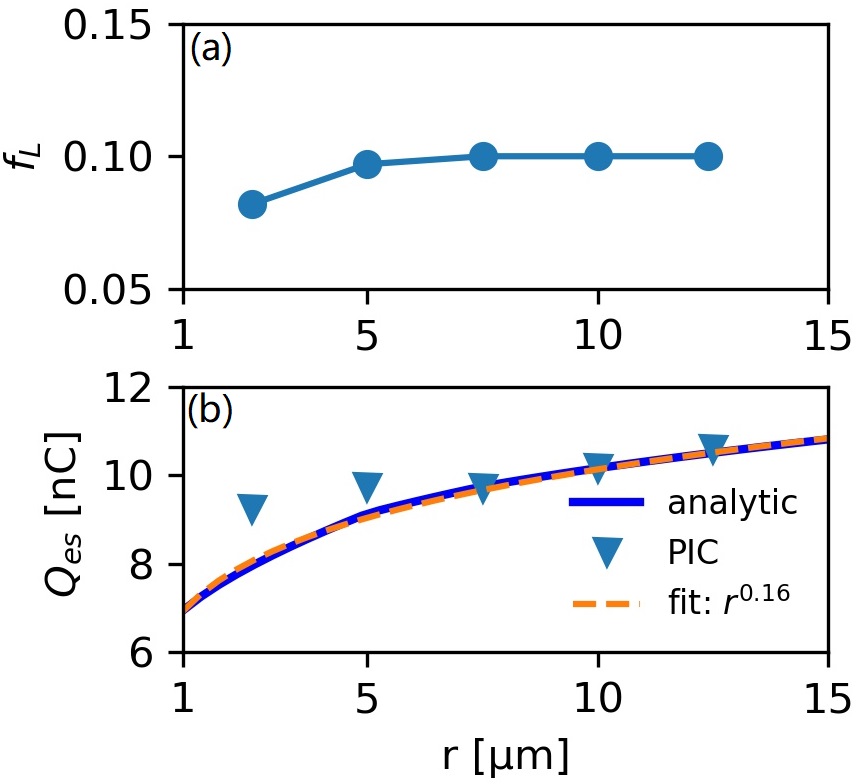} 
\caption{(a) The energy apportion rate obtained from the 2D PIC simulations. (b) The escaping charges are calculated by the PIC simulations and from the analytic formula. The dashed line shows a curve fitting to the analytic results.}
\label{fig:qes_scaling} 
\end{figure}

From Eq.~\eqref{eq:Wfunction}, it is shown that 
\begin{equation}
    \eta(L_0,r) \propto L_0C(r)W(f_w(L_0,C(r)))\approx
    L_0C(r)W(f_w(L_0)),
    \label{eq:law_app}
\end{equation}
where $f_w$ is the collection of term inside the Lambert W function as shown in Eq.~\eqref{eq:Wfunction}, $L_0$ is the effective length of the capacitance and $C_{eff}=L_0C(r)$. The approximation in Eq.~\eqref{eq:law_app} is made due to the fact that $r$ is only a small perturbation compared to $L_0$. Thus, the trend of $\eta(L_0,r)$ varied with target radius is decoupled from target length $L_0$, enabling us to use two-dimensional particle-in-cell (PIC) simulations to verify this result. To assess the influence of $f_L$ on escape charges, 2D PIC simulations were performed using the PIConGPU platform \cite{PIConGPU2013}. In these 2D PIC simulations of hydrogen, the simulation box measures \SI{40}{\um} in both the $x$- and $y$-directions, with a spatial resolution of \SI{0.8}{\um}/96 to resolve the plasma wave. The target is positioned at the center of the simulation box with a specific radius representing the solid hydrogen jet, and no preplasma scale length is included. Figure~\ref{fig:qes_scaling} shows the simulation results. In Figure~\ref{fig:qes_scaling}(a), it is evident that $f_L$ is not constant; its value is 0.082, 0.097, and 0.1 for target radii of \SI{2.5}{\um}, \SI{5}{\um}, and \SI{7.5}{\um}, respectively, with saturation occurring for radii larger than \SI{7.5}{\um}.

The theoretical prediction for the escaping charges, based on Equation~\eqref{eq:Wfunction}, is shown as a solid line in Figure~\ref{fig:qes_scaling}(b). In this calculation, the values of $f_L$ obtained from the PIC simulations were used to adjust the computation of $Q_{es}(r)$. The dotted data in Figure~\ref{fig:qes_scaling}(b) correspond to the escaping charges from the PIC simulations, normalized for comparison with the analytical results. The analytic model agrees well with the simulation data for radii exceeding \SI{5}{\um}, with a deviation of approximately 10\% observed at \SI{2.5}{\um}. Both the simulations and analytical results indicate that the target radius has only a minor effect on the escaping charges. An exponential function of the form $r^x$ was used to fit the data (dashed line in Figure~\ref{fig:qes_scaling}(b)), yielding an exponent of 0.16. This weak dependence of the escaping charges on the radius consequently reduces the scaling effect of the target radius on the surface return current.

\end{appendix}

\clearpage
\bibliography{ref}

\end{document}